\documentclass[12pt]{iopart}
\usepackage{latexsym}
\usepackage{graphicx}
\usepackage{amssymb}
\usepackage{amsthm}
\usepackage{amsfonts}
\usepackage{bbm}

\newcommand{\un}[1]{{\underline{#1}}}

\newcommand{\bra}[1]{{\langle #1 \vert}}

\newcommand{\ket}[1]{{\vert #1 \rangle}}

\newcommand{\ZZ}{\mathbb{Z}}

\newcommand{\be}{\begin{equation}}
\newcommand{\ee}{\end{equation}}
\newcommand{\bea}{\begin{eqnarray}}
\newcommand{\eea}{\end{eqnarray}}

\def\one{\mathbbm{1}}

\begin{document}

\title[On two non-ergodic reversible cellular automata, one classical, the other quantum]{On two non-ergodic reversible cellular automata, one classical, the other quantum}

\author{Toma\v z Prosen}

\address{Faculty of Mathematics and Physics, University of Ljubljana, Jadranska 19, SI-1000 Ljubljana, Slovenia}
\ead{tomaz.prosen@fmf.uni-lj.si}

\begin{abstract}
We propose and discuss two variants of kinetic particle models --- cellular automata in 1+1 dimensions, which have some appeal due to their simplicity and intriguing properties which could warrant further research and applications.
The first model is a deterministic and reversible automaton describing two species of quasiparticles: stable massless \emph{matter} particles moving with velocity $\pm 1$ and unstable, standing (zero velocity) \emph{field} particles. 
We discuss two distinct continuity equations for three conserved charges of the model. While the first two charges and the corresponding currents have support three (3) lattice sites and represent a lattice analogue of conserved energy-momentum tensor, we find an additional conserved charge and current with support of nine (9) sites, implying non-ergodic behaviour and potentially signalling integrability of the model with a highly nested R-matrix structure. 
The second model represents a quantum (or stochastic) deformation of a recently introduced and studied \emph{charged hardpoint lattice gas}, where particles of different binary charge ($\pm 1$) and binary velocity ($\pm 1$)  can nontrivially mix upon elastic collisional scattering. We show that while the unitary evolution rule of this model does not satisfy the full Yang-Baxter equation, it still satisfies an intriguing related identity which gives birth to an infinite set of local conserved operators, the-so-called glider operators.
\\\\
To Prof. Giulio Casati, on the occasion of his 80th birthday
\end{abstract}
  
\section{Introduction}

This article is dedicated to my friend and mentor, Giulio Casati. For that reason I will also take the liberty to write the paper from a very personal perspective. One of the most important things that Giulio taught me, was to passionately appreciate extremely 
simple models of dynamics. Constructing and solving simple models of (lately most often, many-body) dynamics with carefully chosen physical properties has thus become my personal obsession throughout my career. The other thing which I owe Giulio was a pragmatic but deep appreciation of ergodic theory. Together we have been responsible for a few endeavours in experimental mathematics, which sometimes rose attention of experts. For instance, we pointed out intriguing ergodic properties of classical and quantum polygonal billiards and related non-hyperbolic and non-integrable dynamical systems ~\cite{CP1999,CP2000,WCP2014,LCP2022}.

Crudely speaking, ergodic theory~\cite{CFS82} divides dynamical systems into ergodic and non-ergodic ones. The former are characterized by the property that time averages of (relevant) observables can be computed in terms of ensemble average over the entire space of states (phase space), while the latter exhibit memory of the initial condition in typical (say, physically relevant) observables.
Additionally, ergodic dynamics can have various degree of dynamical complexity or chaos, while the non-ergodic ones, can conform to some algebraic tools of exact solvability, or integrability. However, it is not clear to what extent breaking of ergodicity is connected to any form of integrability as it can precisely be defined in terms of Lax pairs (in classical, deterministic systems) or Yang-Baxter equation (in quantum or stochastic setting). In physics literature, these questions have been extensively discussed recently in the context of many-body dynamics, where mathematics of ergodic theory is much less developed, say on the lattice and having local interactions, both in the classical and quantum realm (see e.g. some recent reviews~\cite{review1,review2,review3,review4}). Besides integrability, many different forms of ergodicity breaking have been suggested, such as phase space or Hilbert space fragmentation due to various forms of kinetic constraints, the so-called many-body localization due to static disorder, many-body scarring due to hidden weakly broken non-abelian symmetries, etc.

In this paper we propose two very simple many-body locally interacting dynamical systems defined on a discrete 1+1 space-time lattice.
The first model, which we call a matter-field automation for reasons that will hopefully appear clear to the reader, is an example of a deterministic particle kinematics with two distinct types of degrees of freedom (matter\&field), which is non-ergodic for a nontrivial reason: existence of a highly nontrivial but local conserved charge. Inspired by brute-force empirical computer-algebra calculations, we suggest that the model may be integrable, but if it is, the corresponding R-matrix should be highly nontrivial.
The second model, which is a quantum deformation of the previously studied charged hardpoint lattice gas~\cite{medenjak17,slava,krajnik22},
obeys a remarkable reduced Yang-Baxter-like identity for a non-abelian spectral parameter. This again ensures manifest ergodicity breaking in the model through an existence of an infinite set of local translationally invariant conserved operators, the so-called gliders.

\section{Matter-Field Automaton}

\subsection{Definition of the automaton}

Let us define a deterministic reversible cellular automaton on $\{0,1\}^\ZZ \ni \un{s}$ as follows.
Even lattice sites correspond to {\em matter} variable $s_{2x}\in\{0,1\}$ while odd lattice sites correspond to {\em field} variable $s_{2x+1}\in\{0,1\}$, $x\in\ZZ$.
The instantaneous system configuration is thus specified by an infinite binary sequence $\un{s}$, on which we specify deterministic dynamics.

\begin{figure}
\centering	
\vspace{-1mm}
\includegraphics[width=0.68\columnwidth]{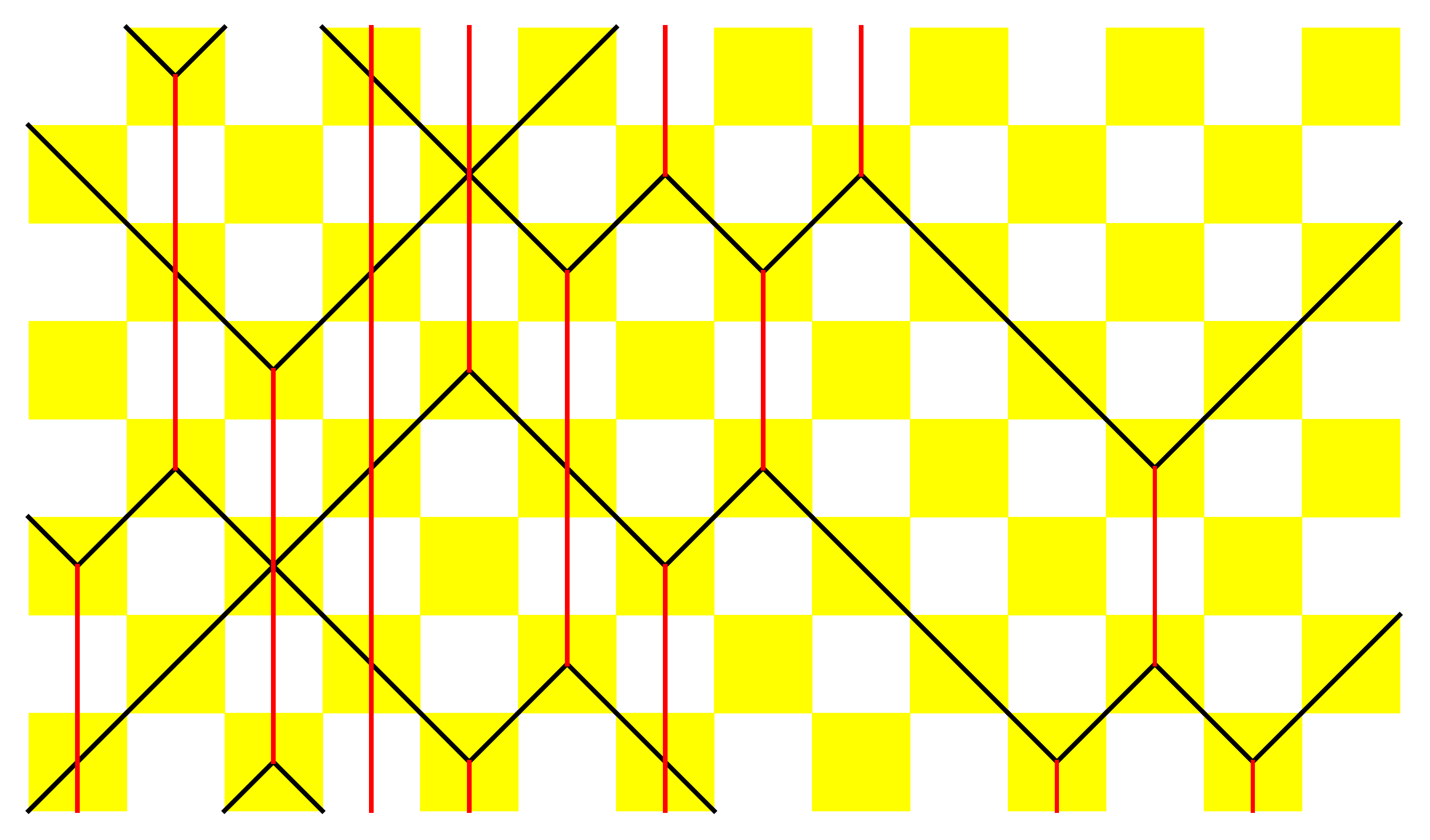}
\vspace{-1mm}
\caption{Example of an automaton's trajectory (time runs in vertical direction, say upwards). Yellow squares denote application of the local rule, or equivalently, of the three-bit permutation gate $Y$. Black/red lines denote the worldlines of matter/field quasiparticles.}
\label{fig:1}
\end{figure}

The dynamics is defined in a staggered fashion, in even and odd time layers.
Specifically, for so-called \emph{even} time steps we update the matter-field-matter triples $(s_{4x},s_{4x+1},s_{4x+2})$, for all $x\in\ZZ$, with a deterministic, reversible rule 
$(ss's'')\longrightarrow(rr'r'')$, while for \emph{odd} time-steps, the triples $(s_{4x-2},s_{4x-1},s_{4x})$, are updated by the same rule. The rule is specified as
\bea
& (000) \longleftrightarrow (000), \nonumber\\
& (001) \longleftrightarrow  (100), \nonumber \\
& (011) \longleftrightarrow (110), \label{rules1} \\
& (101) \longleftrightarrow (010), \nonumber\\
& (111) \longleftrightarrow (111), \nonumber
\eea
or, graphically and perhaps more intuitively, by the following diagrams
\begin{center}
\vspace{3.45mm}

\hbox{\hspace{3.2cm} 0 0 0 \hspace{1.23mm}  1 0 0 \hspace{1.23mm}  1 0 1  \hspace{1.23mm}  1 1 0 \hspace{1.23mm}  0 0 1  \hspace{1.23mm}  0 1 0 \hspace{1.23mm}   0 1 1 \hspace{1.23mm}   1 1 1}

\vspace{2mm}

\hbox{
\hspace{3.03cm} \includegraphics[width=9.5mm]{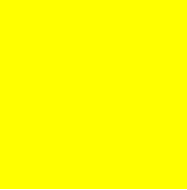}\hspace{3.45mm}\includegraphics[width=9.5mm]{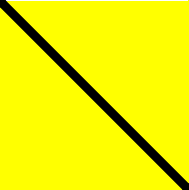}\hspace{3.45mm}\includegraphics[width=9.5mm]{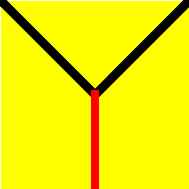}\hspace{3.45mm}\includegraphics[width=9.5mm]{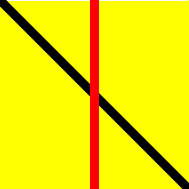}\hspace{3.45mm}\includegraphics[width=9.5mm]{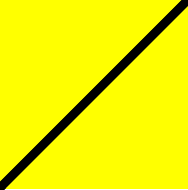}\hspace{3.45mm}\includegraphics[width=9.5mm]{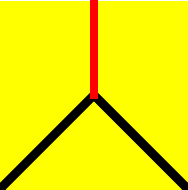}\hspace{3.45mm}\includegraphics[width=9.5mm]{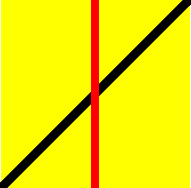}\hspace{3.45mm}\includegraphics[width=9.5mm]{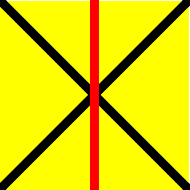}
}

\hbox{\hspace{3.2cm} 0 0 0 \hspace{1.23mm}  0 0 1 \hspace{1.23mm}  0 1 0  \hspace{1.23mm}  0 1 1 \hspace{1.23mm}  1 0 0  \hspace{1.23mm}  1 0 1 \hspace{1.23mm}   1 1 0 \hspace{1.23mm}   1 1 1}

\end{center}
which indicates mappings of all eight lower matter-field-matter triples to corresponding upper matter-field-matter triples, or vice versa due to time reversal symmetry. The black and red 
lines thus represent the worldlines of matter and field particles, respectively. In Fig.~\ref{fig:1} we show an example of a piece of a many-body trajectory $\{s_x^t; x,t\in\ZZ\}$, where time $t$ runs in vertical direction (upwards). The construction of the trajectory can be understood as a checkerboard tiling of the rule's yellow squares which is uniquely prescribed by the lowest two lines of tiles (the initial condition).

Note that this dynamical system has a rich set of three independent $\ZZ^2$ symmetries, specifically C,P, and T: 
\begin{enumerate}
\item ``Charge'' conjugation C, $s\to 1-s$, i.e., if $s^t_x$ is a valid trajectory, then $1-s^t_x$ is a valid trajectory,
\item Parity P, $x\to-x$, i.e., if $s^t_x$ is a valid trajectory, then $s^t_{-x}$ is a valid trajectory, 
\item Time-reversal T, $t\to -t$, i.e., if $s^t_x$ is a valid trajectory, then $s^{-t}_{x}$ is a valid trajectory.
\end{enumerate}

\begin{figure}
\centering	
\vspace{-1mm}
\includegraphics[width=\columnwidth]{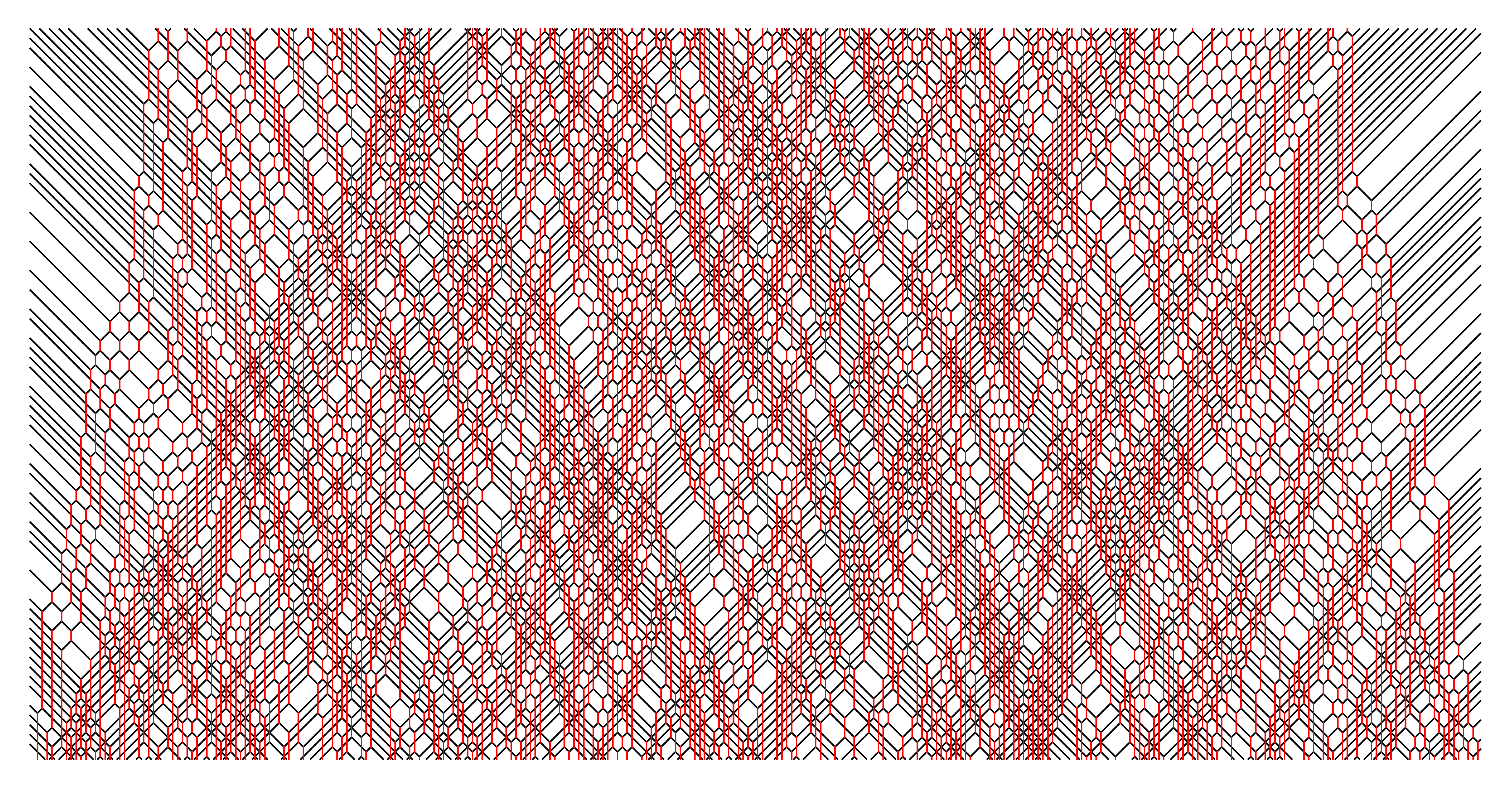}
\vspace{-1mm}
\caption{Example of a typical trajectory in a maximum entropy initial state, where $s_x^{t=0}$ are sampled randomly and uniformly in $\{0,1\}$, for a section of $0\le x\le 600$ and duration
$t_{\rm max}=150$, $0\le t\le t_{\rm max}$. Note the evaporation of matter particles from the sides, and reduction of the field intensity, which is due to initial vacuum in the complement region 
$s^{t=0}_{x'} = 0$, $x'<0$ or $x'>600$.}
\label{fig:2}
\end{figure}

\begin{figure}
\centering	
\vspace{-1mm}
\includegraphics[width=\columnwidth]{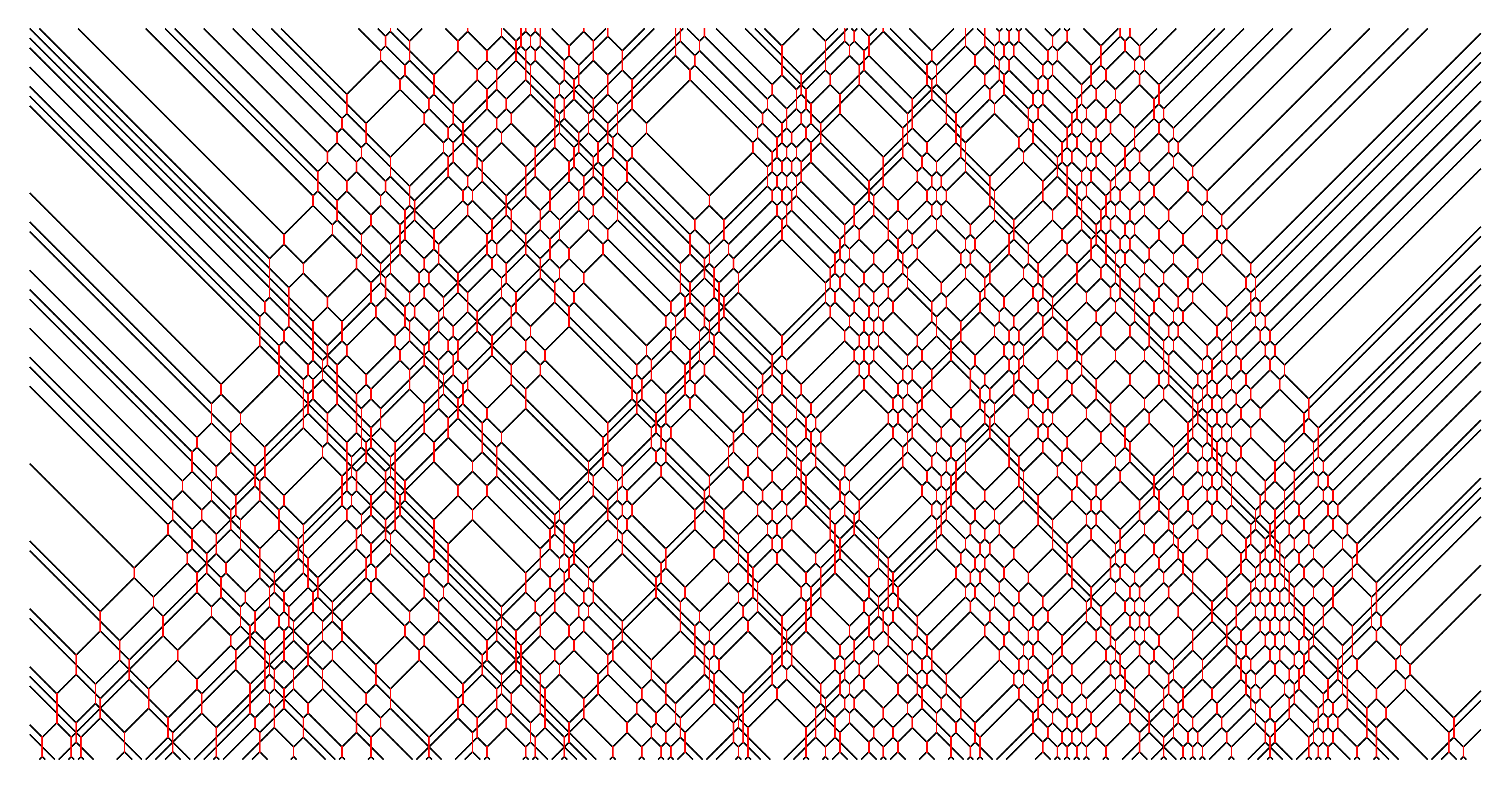}
\vspace{-1mm}
\caption{Example of a typical trajectory from initial state where the matter is in maximal entropy state, i.e. $s_{2x}$ are sampled randomly and uniformly for $0\le 2x\le 600$, 
and there is \emph{no} field, i.e. $s_{2x+1}\equiv 0$, of duration
$t_{\rm max}=150$, $0\le t\le t_{\rm max}$.}
\label{fig:3}
\end{figure}

\begin{figure}
\centering	
\vspace{-1mm}
\includegraphics[width=\columnwidth]{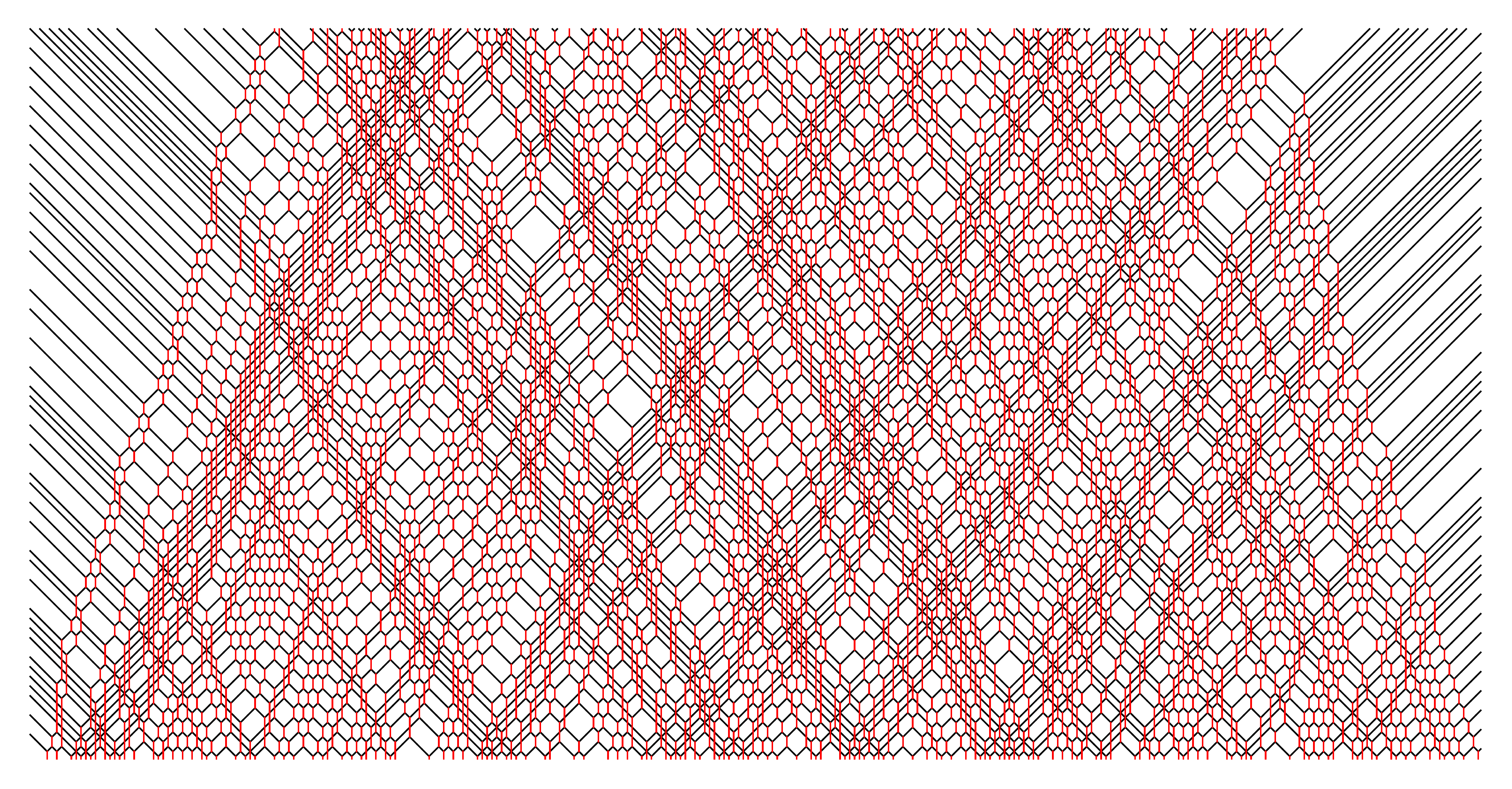}
\vspace{-1mm}
\caption{Example of a typical trajectory from initial state where the field is in maximal entropy state, i.e. $s_{2x+1}$ are sampled randomly and uniformly for $0\le 2x+1\le 600$, 
and there is \emph{no} matter, i.e. $s_{2x}\equiv 0$, of duration
$t_{\rm max}=150$, $0\le t\le t_{\rm max}$.}
\label{fig:4}
\end{figure}

\begin{figure}
\centering	
\vspace{-1mm}
\includegraphics[width=\columnwidth]{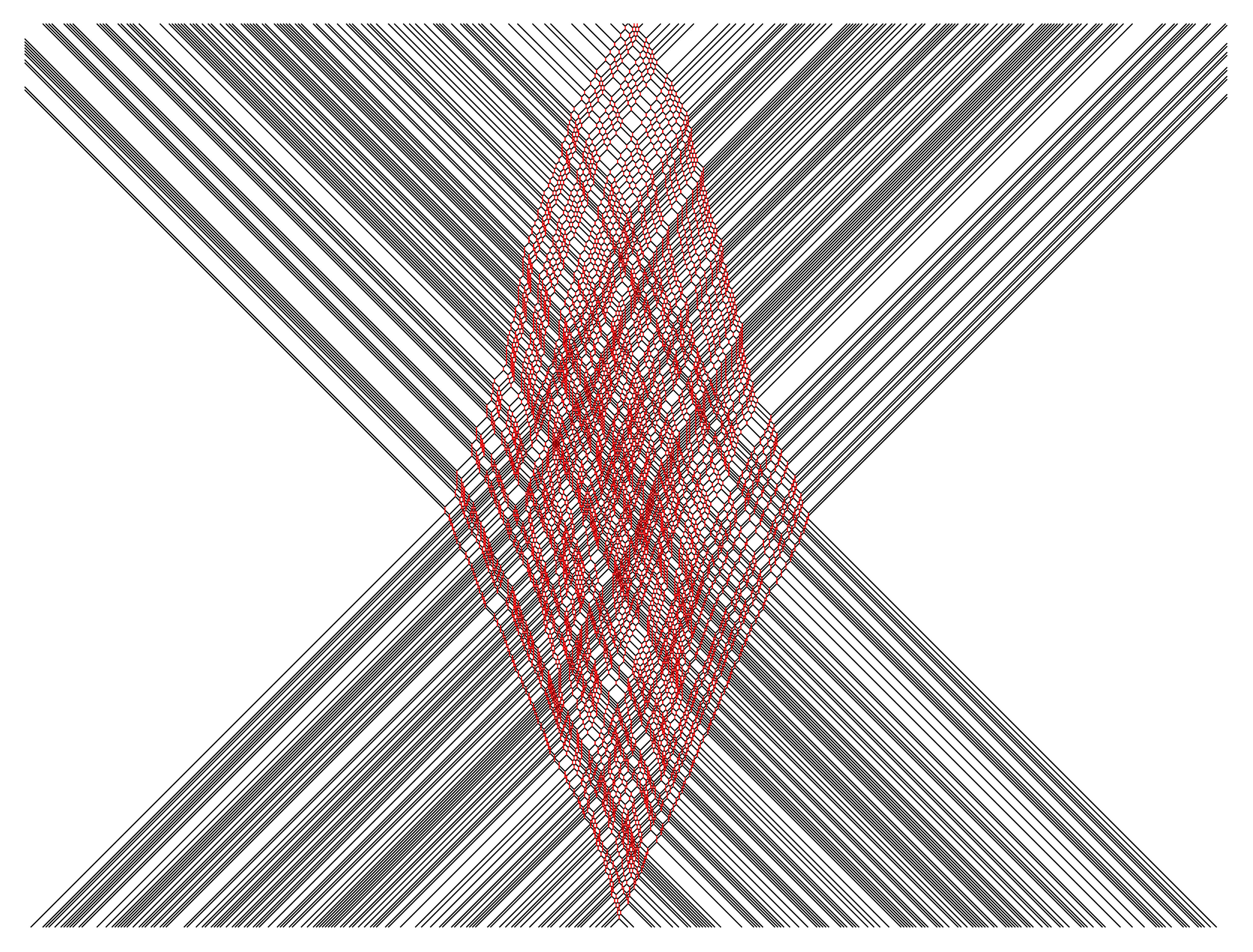}
\vspace{-1mm}
\caption{Head-on collision of a random (maximum entropy) sample of left movers with random sample of right movers. The total width of each initial stream of left/right movers is $800$ and duration is $t_{\rm max}=600$.}
\label{fig:5}
\end{figure}

\begin{figure}
\centering	
\vspace{-1mm}
\includegraphics[width=\columnwidth]{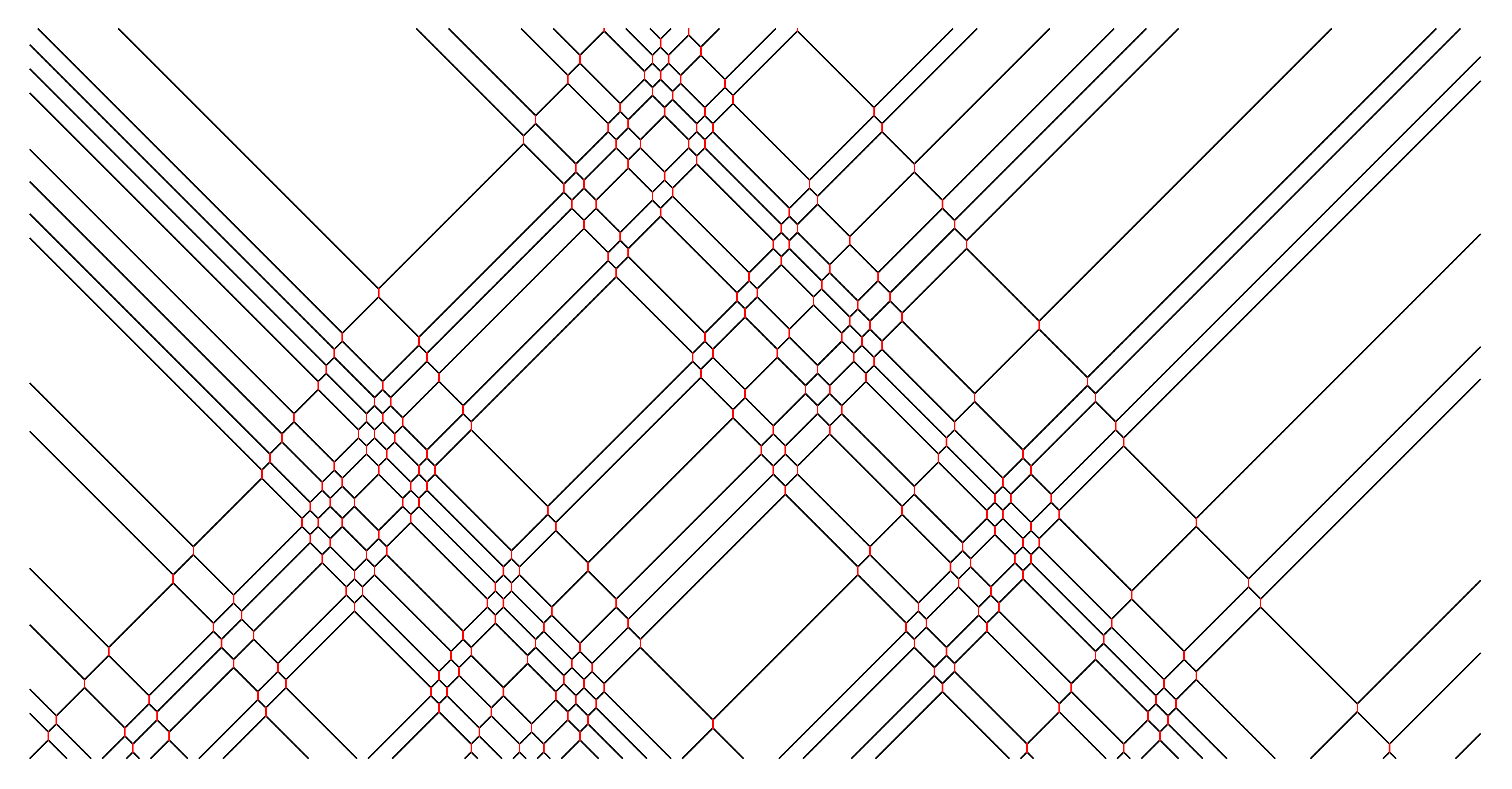}
\vspace{-1mm}
\caption{Trajectory starting from a random initial condition where only $s_{6x}$ ($0\le x \le 180$) are chosen random, uniform in $\{0,1\}$, while all other $s_{x'}=0$.
Such trajectory features only maximally short-lived field particles which provide `unit' phase shifts for left/right movers upon scattering and corresponds to dynamics of negative-length hard-rods, like
 e.g. in the Rule 54 system~\cite{rule54review}.}
\label{fig:6}
\end{figure}

In Figs.~\ref{fig:2}-\ref{fig:6} we depict some interesting characteristic trajectories of the matter-field automaton.
In Fig.~\ref{fig:2} we prepare a maximum entropy initial state in a finite box of size $L$, where both field and matter are sampled randomly and uniformly in the set $\{0,1\}$ and plot an emerging deterministic trajectory. Since the initial condition vanishes outside the box, $s^{t=0}_x=0$, for $x<0$ or $x>L$, we see a slow `evaporation' of matter particles from the edges of the box.
In Fig.~\ref{fig:3} we plot a similar trajectory, where initially we have only matter (at maximum entropy) and no field, while in Fig.~\ref{fig:4} we plot the reverse situation where initially there is only the field at maximum entropy and no matter. The fact that spatio-temporal patterns of these trajectory seem qualitatively distinct depending on the initial condition, and in particular the relative ratio of field versus matter drastically differs, suggests that dynamics may be non-ergodic. We shall establish this more precisely below by constructing non-trivial algebraic conservation laws.  While previous plots represent cartoons of equilibrium dynamics, we plot a typical trajectory representing far-from-equilibrium dynamics in Fig.~\ref{fig:5}: a head on collision of two streams of matter (with no field in the initial data), which experience a non-trivial scattering process via formation of a complex field pattern, finally resulting in two oppositely moving scattered streams of matter.

Finally, we demonstrate that matter-field system can be understood as a very general automaton which can also mimic the dynamics of popular Rule 54 reversible cellular automation~\cite{rule54review} for a dynamically closed (invariant) subset of initial configurations. More generally, we can show it can simulate dynamics for classes of negative-length hard-rod systems with commensurate initial data. For instance, if we take initial condition $\un{s}$ where only every sixth site can be occupied, i.e. $s_{x}\equiv 0$ for $x\neq 0\pmod{6}$, then one can show that no field worldline can be crossed by another matter particle, and hence each field particle decays at immediate next time step (the field is maximally short-lived).
This means that scattering of matter particles always experiences a fixed phase-shift, which makes dynamics qualitatively similar (yet not precisely equivalent) to Rule 54 on a certain reduced lattice (see Fig.~\ref{fig:6} for a demonstration). In Rule 54 dynamics, for instance, the minimal distance between movers/kinks (=2) is twice the time-tag in kink-scattering (=1), while in this case, 
the minimal distance between (left/right) movers (=6) is three times a scattering time-lag (=2), i.e. free field life-time.
Similar commensurate negative-length hard-rod dynamics is obtained for the subset of initial data $\{\un{s}\}$ of the form 
$s_{x'}=0$, $x'\neq 0 \pmod{2(2k+1)}$, for any $k=1,2\ldots$, where the ratio between minimal distance between parallel kinks to scattering-lag is $2k+1$.

\subsection{Time evolution over abelian algebra of local observables}
Let us now reinterpret dynamics of matter-field automaton in terms of an abelian $C^*$-algebraic dynamical system, and then study its conservation laws. We begin by defining a quasilocal algebra of observables, i.e. real functions over space of configurations $\{0,1\}^\ZZ$. For more details on a fully analogous but more precise construction in a related model of Rule 54 
reversible cellular automaton see Ref.~\cite{CMP}.
Let $[0]_{2x}$, $[1]_{2x}$ define a local basis of observables, checking if $s_{2x}=0$ (or $1)$ for the matter degrees of freedom, and correspondingly $[0]_{2x+1},[1]_{2x+1}$ for the
field:
\be
[r]_x (\un{s}) = \delta_{r,s_x}.
\ee
We have $[0]_x+[1]_x=\one$ (unit element in the algebra of observables), and denote $[\bullet]_x = [1]_x-[0]_x$, for $x$ either even or odd.
we note that $[0]_x,[1]_x$ (or $\one,[\bullet]_x$) generate the entire (quasi)local algebra under multiplication, addition (and closure). We denote local observables with larger support as
$[s_0,s_1,\ldots,s_r]_x = [s_0]_x [s_1]_{x+1} \cdots [s_r]_{x+r}$.

The fundamental three site operator -- local propagator -- in the algebra of observables implementing the rules
\bea
& [000]_{2x} \longleftrightarrow[000]_{2x}, \nonumber \\
& [001]_{2x} \longleftrightarrow  [100]_{2x}, \nonumber \\
& [011]_{2x} \longleftrightarrow [110]_{2x}, \\
& [101]_{2x} \longleftrightarrow [010]_{2x}, \nonumber \\
& [111]_{2x} \longleftrightarrow [111]_{2x}, \nonumber
\eea
is encoded in terms of $8\times 8$ permutation matrix
\be
Y=
\left(
\begin{array}{cccccccc}
 1 & 0 & 0 & 0 & 0 & 0 & 0 & 0 \\
 0 & 0 & 0 & 0 & 1 & 0 & 0 & 0 \\
 0 & 0 & 0 & 0 & 0 & 1 & 0 & 0 \\
 0 & 0 & 0 & 0 & 0 & 0 & 1 & 0 \\
 0 & 1 & 0 & 0 & 0 & 0 & 0 & 0 \\
 0 & 0 & 1 & 0 & 0 & 0 & 0 & 0 \\
 0 & 0 & 0 & 1 & 0 & 0 & 0 & 0 \\
 0 & 0 & 0 & 0 & 0 & 0 & 0 & 1 \\
\end{array}
\right)\,.
\label{Y}
\ee
The full time step of dynamical automorphism on the algebra of observables $a^{t}(\un{s})=a(\un{s}^{-t})$, 
\be
a^{2(t+1)}=U a^{2t},\qquad
a^{2t+1}= U_{\rm e} a^{2t},\quad
a^{2t+2}= U_{\rm o} a^{2t+1},
\ee 
is constructed as follows
\be
U = U_{\rm o} U_{\rm e}\,,
\ee
where even and odd time steps are generated as
\bea
U_{\rm e} = \prod_x Y_{4x,4x+1,4x+2},\quad
U_{\rm o} = \prod_x Y_{4x-2,4x-1,4x}\,,
\eea
where three indices denote the positions in the string, or in tensor product of local algebras ($\bigotimes_x {\rm span}\{ [0]_x,[1]_x\}$), where the three site operator (\ref{Y}) acts nontrivially.

\subsection{Conservation laws}

Let $a_{[x,y]}$ be any local observable supported in the interval of sites $[x,y]\subset \ZZ$.
We define a shift automorphism $\Pi$
\be
\Pi a_{[x,y]} = a_{[x+2,y+2]}\,,
\ee
which shifts an observable by `one lattice unit', so that the identity of matter/field sites is preserved.
Note that
\be
\Pi U^{\rm e} = U^{\rm o} \Pi,  \quad \Pi U^{\rm o} = U^{\rm e} \Pi,  \quad [\Pi^2, U^{\rm e}] = 0,\quad [\Pi^2, U^{\rm o}]=0.
\ee
A local observable $q$ is called a conserved charge, if together with an appropriate conserved current $p$, satisfies a conservation law (space-time discrete continuity equation):
\be
U_{\rm e} q - U_{\rm o} q + \Pi p - \Pi^{-1} p = 0.
\label{CL}
\ee
Then, extensive local observable is globally conserved
\be
Q = \sum_{x\in\ZZ} q_x,
\quad
q_x:=\Pi^{2x} q,\qquad U Q = Q.
\ee
Limiting to local observables on clusters of \emph{five} sites (beginning end ending with a matter site), covering one 'unit cell' (periodicity block of 4 sites) of the automaton, we find (using exact computer algebra) exactly two conservation laws
\bea
q^{(1)}_x &=& [\bullet]_{4x} - [\bullet]_{4x+2}, \label{eq:cl1}\\
p^{(1)}_x &=& \frac{1}{2} [\bullet]_{4x} +  \frac{1}{2}[\bullet]_{4x+1} +
\frac{1}{2} [\bullet]_{4x+2} +  \frac{1}{2} [\bullet]_{4x} [\bullet]_{4x+1}[\bullet]_{4x+2},  \label{eq:cl2} \\
 q^{(2)}_x &=& [\bullet]_{4x+1} + [\bullet]_{4x+2} + [\bullet]_{4x+3},\\
p_x^{(2)} &=& - [\bullet]_{4x+2}, 
\eea
with local densities and current supported on up to three (3) consecutive sites.
$q^{(1)}_x$ can be interpreted as the {\em momentum density} while $q^{(2)}_x$ is the analog of the {\em energy density}.
Indeed, writing $P = \sum_x q^{(1)}_x$, $E=\sum_x q^{(2)}_x$ we have an intuitive expressions for total momentum and total energy of a configuration:
\be
P(\un{s}) = \sum_x (-1)^x s_{2x}\,, \quad
E(\un{s}) = \sum_x ( s_{2x} + 2 s_{2x+1})\,,
\label{EP}
\ee
meaning that the field particle caries no momentum and has an energy content (rest energy) equal to the energy of two matter particles.

If we then increase the support size to clusers of \emph{nine} ($2\times 4+1$) consecutive sites, we obtain precisely one new additional conservation law:\\
(using compact notation: $[\circ]\equiv \one$, $[r_0r_1r_2\ldots]_{4x} \equiv [r_0r_1r_2\ldots] $)

\bea
&\!\!\!\!\!\!\!\!\!\!\!\!\!\!\!\!\!\!\!\!\!\!\!\!\!\!\!\!\!\!\!\!\!q^{(3)} = 2 [\circ\circ\circ\circ\circ\circ\bullet\circ\bullet]-2 [\circ\circ\circ\circ\circ\circ\bullet\bullet\circ]-4 [\circ\circ\circ\circ\circ\bullet\circ\circ\bullet]-2
   [\circ\circ\circ\circ\circ\bullet\bullet\circ\circ]
    \nonumber\\ &\!\!\!\!\!\!\!\!\!\!\!\!\!\!\!\!\!\!\!\!\!\!\!\!
   +2 [\circ\circ\circ\circ\circ\bullet\bullet\bullet\bullet]-8 [\circ\circ\circ\circ\bullet\circ\circ\circ\bullet]-4
   [\circ\circ\circ\circ\bullet\circ\circ\bullet\circ]+2 [\circ\circ\circ\circ\bullet\circ\bullet\circ\circ]
     \nonumber\\ &\!\!\!\!\!\!\!\!\!\!\!\!\!\!\!\!\!\!\!\!\!\!\!\!
   -2 [\circ\circ\circ\circ\bullet\circ\bullet\bullet\bullet]+4
   [\circ\circ\circ\circ\bullet\bullet\circ\bullet\bullet]-2 [\circ\circ\circ\circ\bullet\bullet\bullet\circ\bullet]+2 [\circ\circ\circ\circ\bullet\bullet\bullet\bullet\circ]
     \nonumber\\ &\!\!\!\!\!\!\!\!\!\!\!\!\!\!\!\!\!\!\!\!\!\!\!\!
   -2[\circ\circ\circ\bullet\circ\circ\circ\circ\bullet]-2 [\circ\circ\circ\bullet\circ\circ\circ\bullet\circ]+2 [\circ\circ\circ\bullet\circ\circ\bullet\circ\circ]-2
   [\circ\circ\circ\bullet\circ\circ\bullet\bullet\bullet]
     \nonumber\\ &\!\!\!\!\!\!\!\!\!\!\!\!\!\!\!\!\!\!\!\!\!\!\!\!
   -4 [\circ\circ\circ\bullet\circ\bullet\circ\circ\circ]+2 [\circ\circ\circ\bullet\bullet\bullet\circ\circ\bullet]+2
   [\circ\circ\circ\bullet\bullet\bullet\circ\bullet\circ]-2 [\circ\circ\circ\bullet\bullet\bullet\bullet\circ\circ]
     \nonumber\\ &\!\!\!\!\!\!\!\!\!\!\!\!\!\!\!\!\!\!\!\!\!\!\!\!
   +2
   [\circ\circ\circ\bullet\bullet\bullet\bullet\bullet\bullet]+[\circ\circ\bullet\circ\circ\circ\circ\circ\bullet]+[\circ\circ\bullet\circ\circ\circ\circ\bullet\circ]-[\circ\circ\bullet\circ\circ\circ\bullet\circ
   \circ]
     \nonumber\\ &\!\!\!\!\!\!\!\!\!\!\!\!\!\!\!\!\!\!\!\!\!\!\!\!
   +[\circ\circ\bullet\circ\circ\circ\bullet\bullet\bullet]+2
   [\circ\circ\bullet\circ\circ\bullet\circ\circ\circ]-[\circ\circ\bullet\circ\bullet\bullet\circ\circ\bullet]-[\circ\circ\bullet\circ\bullet\bullet\circ\bullet\circ]
    \nonumber\\ &\!\!\!\!\!\!\!\!\!\!\!\!\!\!\!\!\!\!\!\!\!\!\!\!
   +[\circ\circ\bullet\circ\bullet\bullet\bullet\circ
   \circ]-[\circ\circ\bullet\circ\bullet\bullet\bullet\bullet\bullet]
   +[\circ\circ\bullet\bullet\circ\bullet\circ\circ\bullet]+[\circ\circ\bullet\bullet\circ\bullet\circ\bullet\circ]
    \nonumber\\ &\!\!\!\!\!\!\!\!\!\!\!\!\!\!\!\!\!\!\!\!\!\!\!\!
   -[\circ\circ\bullet\bullet\circ\bullet
   \bullet\circ\circ]+[\circ\circ\bullet\bullet\circ\bullet\bullet\bullet\bullet]-[\circ\circ\bullet\bullet\bullet\circ\circ\circ\bullet]-[\circ\circ\bullet\bullet\bullet\circ\circ\bullet\circ]
    \nonumber\\ &\!\!\!\!\!\!\!\!\!\!\!\!\!\!\!\!\!\!\!\!\!\!\!\!
   +[\circ\circ\bullet\bullet
   \bullet\circ\bullet\circ\circ]-[\circ\circ\bullet\bullet\bullet\circ\bullet\bullet\bullet]-2
   [\circ\circ\bullet\bullet\bullet\bullet\circ\circ\circ]-[\circ\bullet\circ\circ\circ\circ\circ\circ\bullet]
    \nonumber\\ &\!\!\!\!\!\!\!\!\!\!\!\!\!\!\!\!\!\!\!\!\!\!\!\!
   -[\circ\bullet\circ\circ\circ\circ\circ\bullet\circ]+[\circ\bullet\circ\circ\circ\circ\bullet\circ
   \circ]-[\circ\bullet\circ\circ\circ\circ\bullet\bullet\bullet]-2
   [\circ\bullet\circ\circ\circ\bullet\circ\circ\circ]
    \nonumber\\ &\!\!\!\!\!\!\!\!\!\!\!\!\!\!\!\!\!\!\!\!\!\!\!\!
   +[\circ\bullet\circ\circ\bullet\bullet\circ\circ\bullet]+[\circ\bullet\circ\circ\bullet\bullet\circ\bullet\circ]-[\circ\bullet\circ\circ\bullet\bullet\bullet\circ
   \circ]+[\circ\bullet\circ\circ\bullet\bullet\bullet\bullet\bullet]
    \nonumber\\ &\!\!\!\!\!\!\!\!\!\!\!\!\!\!\!\!\!\!\!\!\!\!\!\!
   -[\circ\bullet\circ\bullet\circ\bullet\circ\circ\bullet]-[\circ\bullet\circ\bullet\circ\bullet\circ\bullet\circ]+[\circ\bullet\circ\bullet\circ\bullet
   \bullet\circ\circ]-[\circ\bullet\circ\bullet\circ\bullet\bullet\bullet\bullet]
    \nonumber\\ &\!\!\!\!\!\!\!\!\!\!\!\!\!\!\!\!\!\!\!\!\!\!\!\!
   +[\circ\bullet\circ\bullet\bullet\circ\circ\circ\bullet]+[\circ\bullet\circ\bullet\bullet\circ\circ\bullet\circ]-[\circ\bullet\circ\bullet
   \bullet\circ\bullet\circ\circ]+[\circ\bullet\circ\bullet\bullet\circ\bullet\bullet\bullet]
    \nonumber\\ &\!\!\!\!\!\!\!\!\!\!\!\!\!\!\!\!\!\!\!\!\!\!\!\!
   +2
   [\circ\bullet\circ\bullet\bullet\bullet\circ\circ\circ]-[\bullet\circ\circ\circ\circ\circ\circ\circ\bullet]-[\bullet\circ\circ\circ\circ\circ\circ\bullet\circ]+[\bullet\circ\circ\circ\circ\circ\bullet\circ
   \circ]
    \nonumber\\ &\!\!\!\!\!\!\!\!\!\!\!\!\!\!\!\!\!\!\!\!\!\!\!\!
   -[\bullet\circ\circ\circ\circ\circ\bullet\bullet\bullet]-2
   [\bullet\circ\circ\circ\circ\bullet\circ\circ\circ]+[\bullet\circ\circ\circ\bullet\bullet\circ\circ\bullet]+[\bullet\circ\circ\circ\bullet\bullet\circ\bullet\circ]
    \nonumber\\ &\!\!\!\!\!\!\!\!\!\!\!\!\!\!\!\!\!\!\!\!\!\!\!\!
   -[\bullet\circ\circ\circ\bullet\bullet\bullet\circ
   \circ]+[\bullet\circ\circ\circ\bullet\bullet\bullet\bullet\bullet]-[\bullet\circ\circ\bullet\circ\bullet\circ\circ\bullet]-[\bullet\circ\circ\bullet\circ\bullet\circ\bullet\circ]
    \nonumber\\ &\!\!\!\!\!\!\!\!\!\!\!\!\!\!\!\!\!\!\!\!\!\!\!\!
   +[\bullet\circ\circ\bullet\circ\bullet
   \bullet\circ\circ]-[\bullet\circ\circ\bullet\circ\bullet\bullet\bullet\bullet]+[\bullet\circ\circ\bullet\bullet\circ\circ\circ\bullet]+[\bullet\circ\circ\bullet\bullet\circ\circ\bullet\circ]
    \nonumber\\ &\!\!\!\!\!\!\!\!\!\!\!\!\!\!\!\!\!\!\!\!\!\!\!\!
   -[\bullet\circ\circ\bullet
   \bullet\circ\bullet\circ\circ]+[\bullet\circ\circ\bullet\bullet\circ\bullet\bullet\bullet]+2
   [\bullet\circ\circ\bullet\bullet\bullet\circ\circ\circ]-[\bullet\bullet\bullet\circ\circ\circ\circ\circ\bullet]
    \nonumber\\ &\!\!\!\!\!\!\!\!\!\!\!\!\!\!\!\!\!\!\!\!\!\!\!\!
   -[\bullet\bullet\bullet\circ\circ\circ\circ\bullet\circ]+[\bullet\bullet\bullet\circ\circ\circ\bullet\circ
   \circ]-[\bullet\bullet\bullet\circ\circ\circ\bullet\bullet\bullet]-2
   [\bullet\bullet\bullet\circ\circ\bullet\circ\circ\circ]
    \nonumber\\ &\!\!\!\!\!\!\!\!\!\!\!\!\!\!\!\!\!\!\!\!\!\!\!\!
   +[\bullet\bullet\bullet\circ\bullet\bullet\circ\circ\bullet]+[\bullet\bullet\bullet\circ\bullet\bullet\circ\bullet\circ]-[\bullet\bullet\bullet\circ\bullet\bullet\bullet\circ
   \circ]+[\bullet\bullet\bullet\circ\bullet\bullet\bullet\bullet\bullet]
    \nonumber\\ &\!\!\!\!\!\!\!\!\!\!\!\!\!\!\!\!\!\!\!\!\!\!\!\!
   -[\bullet\bullet\bullet\bullet\circ\bullet\circ\circ\bullet]
   -[\bullet\bullet\bullet\bullet\circ\bullet\circ\bullet\circ]+[\bullet\bullet\bullet\bullet\circ\bullet
   \bullet\circ\circ]-[\bullet\bullet\bullet\bullet\circ\bullet\bullet\bullet\bullet]
    \nonumber\\ &\!\!\!\!\!\!\!\!\!\!\!\!\!\!\!\!\!\!\!\!\!\!\!\!
   +[\bullet\bullet\bullet\bullet\bullet\circ\circ\circ\bullet]+[\bullet\bullet\bullet\bullet\bullet\circ\circ\bullet\circ]-[\bullet\bullet\bullet\bullet
   \bullet\circ\bullet\circ\circ]+[\bullet\bullet\bullet\bullet\bullet\circ\bullet\bullet\bullet]
    \nonumber\\ &\!\!\!\!\!\!\!\!\!\!\!\!\!\!\!\!\!\!\!\!\!\!\!\!
   +2 [\bullet\bullet\bullet\bullet\bullet\bullet\circ\circ\circ]
\eea
with the current
\bea
&\!\!\!\!\!\!\!\!\!\!\!\!\!\!\!\!\!\!\!\!\!\!\!\!\!\!\!\!\!\!\!\!\!
  p^{(3)}=
2 [\circ\circ\circ\circ\circ\bullet\bullet\circ\circ]-4 [\circ\circ\circ\circ\bullet\circ\circ\circ\bullet]-4 [\circ\circ\circ\circ\bullet\circ\circ\bullet\circ]+2
   [\circ\circ\circ\circ\bullet\circ\bullet\circ\circ]
     \nonumber\\ &\!\!\!\!\!\!\!\!\!\!\!\!\!\!\!\!\!\!\!\!\!\!\!\!
   -4 [\circ\circ\circ\circ\bullet\circ\bullet\bullet\bullet]-2 [\circ\circ\circ\bullet\circ\circ\circ\circ\bullet]-2
   [\circ\circ\circ\bullet\circ\circ\circ\bullet\circ]-2 [\circ\circ\circ\bullet\circ\circ\bullet\bullet\bullet]
     \nonumber\\ &\!\!\!\!\!\!\!\!\!\!\!\!\!\!\!\!\!\!\!\!\!\!\!\!
   +2 [\circ\circ\circ\bullet\bullet\bullet\circ\circ\bullet]+2
   [\circ\circ\circ\bullet\bullet\bullet\circ\bullet\circ]+2
   [\circ\circ\circ\bullet\bullet\bullet\bullet\bullet\bullet]+[\circ\circ\bullet\circ\circ\circ\circ\circ\bullet]
     \nonumber\\ &\!\!\!\!\!\!\!\!\!\!\!\!\!\!\!\!\!\!\!\!\!\!\!\!
   +[\circ\circ\bullet\circ\circ\circ\circ\bullet\circ]+[\circ\circ\bullet\circ\circ\circ\bullet\bullet
   \bullet]-[\circ\circ\bullet\circ\bullet\bullet\circ\circ\bullet]-[\circ\circ\bullet\circ\bullet\bullet\circ\bullet\circ]
     \nonumber\\ &\!\!\!\!\!\!\!\!\!\!\!\!\!\!\!\!\!\!\!\!\!\!\!\!
   -[\circ\circ\bullet\circ\bullet\bullet\bullet\bullet\bullet]+[\circ\circ\bullet\bullet\circ\bullet
   \circ\circ\bullet]+[\circ\circ\bullet\bullet\circ\bullet\circ\bullet\circ]+[\circ\circ\bullet\bullet\circ\bullet\bullet\bullet\bullet]
     \nonumber\\ &\!\!\!\!\!\!\!\!\!\!\!\!\!\!\!\!\!\!\!\!\!\!\!\!
   -[\circ\circ\bullet\bullet\bullet\circ\circ\circ\bullet]-[\circ\circ\bullet\bullet
   \bullet\circ\circ\bullet\circ]-[\circ\circ\bullet\bullet\bullet\circ\bullet\bullet\bullet]-[\circ\bullet\circ\circ\circ\circ\circ\circ\bullet]
     \nonumber\\ &\!\!\!\!\!\!\!\!\!\!\!\!\!\!\!\!\!\!\!\!\!\!\!\!
   -[\circ\bullet\circ\circ\circ\circ\circ\bullet\circ]-[\circ\bullet
   \circ\circ\circ\circ\bullet\bullet\bullet]+[\circ\bullet\circ\circ\bullet\bullet\circ\circ\bullet]+[\circ\bullet\circ\circ\bullet\bullet\circ\bullet\circ]
     \nonumber\\ &\!\!\!\!\!\!\!\!\!\!\!\!\!\!\!\!\!\!\!\!\!\!\!\!
   +[\circ\bullet\circ\circ\bullet\bullet\bullet\bullet\bullet]-[\circ\bullet\circ\bullet\circ\bullet\circ\circ\bullet]-[\circ\bullet\circ\bullet\circ\bullet\circ\bullet\circ]-[\circ\bullet\circ\bullet\circ\bullet\bullet\bullet\bullet]
     \nonumber\\ &\!\!\!\!\!\!\!\!\!\!\!\!\!\!\!\!\!\!\!\!\!\!\!\!
   +[\circ\bullet\circ\bullet\bullet\circ\circ\circ
   \bullet]+[\circ\bullet\circ\bullet\bullet\circ\circ\bullet\circ]+[\circ\bullet\circ\bullet\bullet\circ\bullet\bullet\bullet]-[\bullet\circ\circ\circ\circ\circ\circ\circ\bullet]
     \nonumber\\ &\!\!\!\!\!\!\!\!\!\!\!\!\!\!\!\!\!\!\!\!\!\!\!\!
   -[\bullet\circ\circ\circ\circ\circ
   \circ\bullet\circ]-[\bullet\circ\circ\circ\circ\circ\bullet\bullet\bullet]+[\bullet\circ\circ\circ\bullet\bullet\circ\circ\bullet]+[\bullet\circ\circ\circ\bullet\bullet\circ\bullet\circ]
     \nonumber\\ &\!\!\!\!\!\!\!\!\!\!\!\!\!\!\!\!\!\!\!\!\!\!\!\!
   +[\bullet\circ\circ\circ
   \bullet\bullet\bullet\bullet\bullet]-[\bullet\circ\circ\bullet\circ\bullet\circ\circ\bullet]-[\bullet\circ\circ\bullet\circ\bullet\circ\bullet\circ]-[\bullet\circ\circ\bullet\circ\bullet\bullet\bullet\bullet]
     \nonumber\\ &\!\!\!\!\!\!\!\!\!\!\!\!\!\!\!\!\!\!\!\!\!\!\!\!
   +[\bullet\circ
   \circ\bullet\bullet\circ\circ\circ\bullet]+[\bullet\circ\circ\bullet\bullet\circ\circ\bullet\circ]+[\bullet\circ\circ\bullet\bullet\circ\bullet\bullet\bullet]-[\bullet\bullet\bullet\circ\circ\circ\circ\circ\bullet]
     \nonumber\\ &\!\!\!\!\!\!\!\!\!\!\!\!\!\!\!\!\!\!\!\!\!\!\!\!
   -[
   \bullet\bullet\bullet\circ\circ\circ\circ\bullet\circ]-[\bullet\bullet\bullet\circ\circ\circ\bullet\bullet\bullet]+[\bullet\bullet\bullet\circ\bullet\bullet\circ\circ\bullet]+[\bullet\bullet\bullet\circ\bullet\bullet\circ\bullet\circ
   ]
     \nonumber\\ &\!\!\!\!\!\!\!\!\!\!\!\!\!\!\!\!\!\!\!\!\!\!\!\!
   +[\bullet\bullet\bullet\circ\bullet\bullet\bullet\bullet\bullet]-[\bullet\bullet\bullet\bullet\circ\bullet\circ\circ\bullet]-[\bullet\bullet\bullet\bullet\circ\bullet\circ\bullet\circ]-[\bullet\bullet\bullet\bullet\circ\bullet\bullet
   \bullet\bullet]
     \nonumber\\ &\!\!\!\!\!\!\!\!\!\!\!\!\!\!\!\!\!\!\!\!\!\!\!\!
   +[\bullet\bullet\bullet\bullet\bullet\circ\circ\circ\bullet]+[\bullet\bullet\bullet\bullet\bullet\circ\circ\bullet\circ]+[\bullet\bullet\bullet\bullet\bullet\circ\bullet\bullet\bullet].
\eea
If we furthermore increase to 3-cell cluster of $13=3 \times 4+1$ consecutive sites, we obtain only one additional conservation law
which, however, is trivially related to the previous one:
\be
q^{(4)}_x = \Pi q^{(3)}_x = q^{(3)}_{x+2},\qquad
p^{(4)}_x = -\Pi p^{(3)}_x = -p^{(3)}_{x+2}.
\ee
Of course, it holds in general, that if $(q,p)$ is a local conservation law, then $(\Pi q, -\Pi p)$ is also a conservation law.
Scanning for existence of higher conservation laws with densities supported over four or more unit cells seems impossible with brute-force computer algebra.

Also note that the conservation law property (\ref{CL}) is invariant under local space-time discrete gauge transformation, i.e. taking any local observable $a$
\bea
q &\longrightarrow q + \Pi a - \Pi^{-1} a, \\
p &\longrightarrow p - U_{\rm e} a + U_{\rm o} a.
\eea
In the explicit expressions listed above, the gauge has been fixed by right-alignment of the densities.

One might expect that matter-field automaton hosts an infinite tower of local conserved quantities of increasing support size, however these may be very difficult to find empirically. 
This hypothesis would suggest also that the matter-field automaton may be a completely integrable system in a similar spirit as Rule 54~\cite{balazsrule54}, 
as they also display similar negative-length hard rod dynamics for a class of initial data, but preliminary attempts to find a Yang-Baxter or Lax structure failed.
The other option is that matter-field system has only a few, perhaps finitely many (or an incomplete set) of conserved local charges and would then represent a paradigmatic theory between integrable and ergodic dynamics.

Deterministic cellular automata on discrete phase space (space of configurations) often admit continuous, single (or few) parameter modifications, which render the models quantum or stochastic (depending on the type of deformation which turns a deterministic evolution to a unitary or markov map). For instance, in the definition of matter-field automaton, one may in each step allow for the tripple $(010)$ to remain $(010)$ with probability $\alpha$ and to map to $(101)$ with probability 
 $1-\alpha$. Similarly, $(101)$ remains in $(101)$ with probability $\beta$ and maps to $(010)$ with probability $1-\beta$. All the definitions of the previous section remain valid, except that the permutation matrix $Y$ is now replaced by a stochastic (markov) matrix:
 \be
Y_{\rm stoch}=
\left(
\begin{array}{cccccccc}
 1 & 0 & 0 & 0 & 0 & 0 & 0 & 0 \\
 0 & 0 & 0 & 0 & 1 & 0 & 0 & 0 \\
 0 & 0 & \alpha & 0 & 0 & 1-\beta & 0 & 0 \\
 0 & 0 & 0 & 0 & 0 & 0 & 1 & 0 \\
 0 & 1 & 0 & 0 & 0 & 0 & 0 & 0 \\
 0 & 0 & 1-\alpha & 0 & 0 &  \beta & 0 & 0 \\
 0 & 0 & 0 & 1 & 0 & 0 & 0 & 0 \\
 0 & 0 & 0 & 0 & 0 & 0 & 0 & 1 \\
\end{array}
\right)\,.
\label{Ystoch}
\ee
Interestingly, the first two conservation laws, the momentum and energy (\ref{EP}), see also (\ref{eq:cl1},\ref{eq:cl2}), survive the stochastic deformation, while, as indicated by computer algebra, the third conservation law $(q^{(3)},p^{(3)})$ is \emph{broken}. Hence, the deformation of matter-field automaton likely no longer posses nontrivial ergodicity breaking.

\section{Deformed (quantized) hardpoint lattice gas} 

Nevertheless, we show briefly in this section how continuous deformations of some other reversible deterministic cellular automata can lead to interesting non-ergodic dynamics. 
For this purpose we use the so-called charged hardpoint lattice gas~\cite{medenjak17} (also referred to as XXC model in~\cite{YBmaps} since it is a deterministic limit of models introduced and studied by Maassarani~\cite{XXC}).

In comparison to matter-field automaton, the model has a richer local configuration space, i.e. three states per site, and a simpler local interaction map, applying only to a pair of neighbouring sites instead of three. In physical terms, the model describes dynamics of free point particles which move with speeds $\pm 1$ with initial positions placed at commensurate (multiples of integer) coordinates, but which may carry an internal (say binary) degree of freedom (charge). Denoting the three local states as $s\in\{0,+,-\}$, 0 referred to as a vacancy, the local reversible and deterministic rule, which we apply sequentially to pairs $(2x,2x+1)$ and $(2x+1,2x)$ reads 
\be
(00)\longleftrightarrow(00)\,, \quad(0\alpha)\longleftrightarrow (\alpha0)\,, \quad(\alpha\beta)\longleftrightarrow (\alpha\beta)\,, \quad \alpha,\beta\in\{+,-\}.
\ee
The model allows for a remarkable set of exact results on equilibrium and non-equilibrium dynamics, ranging from dynamical correlations in equilibrium and the transport coefficients~\cite{medenjak17}, to analytic \emph{matrix-product-state} description of time dependent quenches and non-equilibrium steady states in boundary driven setup~\cite{slava}, to recent exact solution of full counting statistics~\cite{krajnik22}.
One may wander if the model still allows for some degree of solvability, if hard-core collision is relaxed to a general stochastic, or unitary scattering
\be
(+-) \longrightarrow u_{11} (+-) +  u_{12} (-+),\qquad 
(-+) \longrightarrow u_{21} (+-) +  u_{22} (-+),
\ee 
where $u$ may be a $2\times 2$ stochastic, or unitary matrix. In the latter case, we think of quantum lattice gas model and quantum superposition states as elaborated precisely below. We note that this system is closely related to a semiclassically quantized sine-Gordon model recently studied in~\cite{kormos}.

\begin{figure}
\centering	
\vspace{-1mm}
\includegraphics[width=0.65\columnwidth]{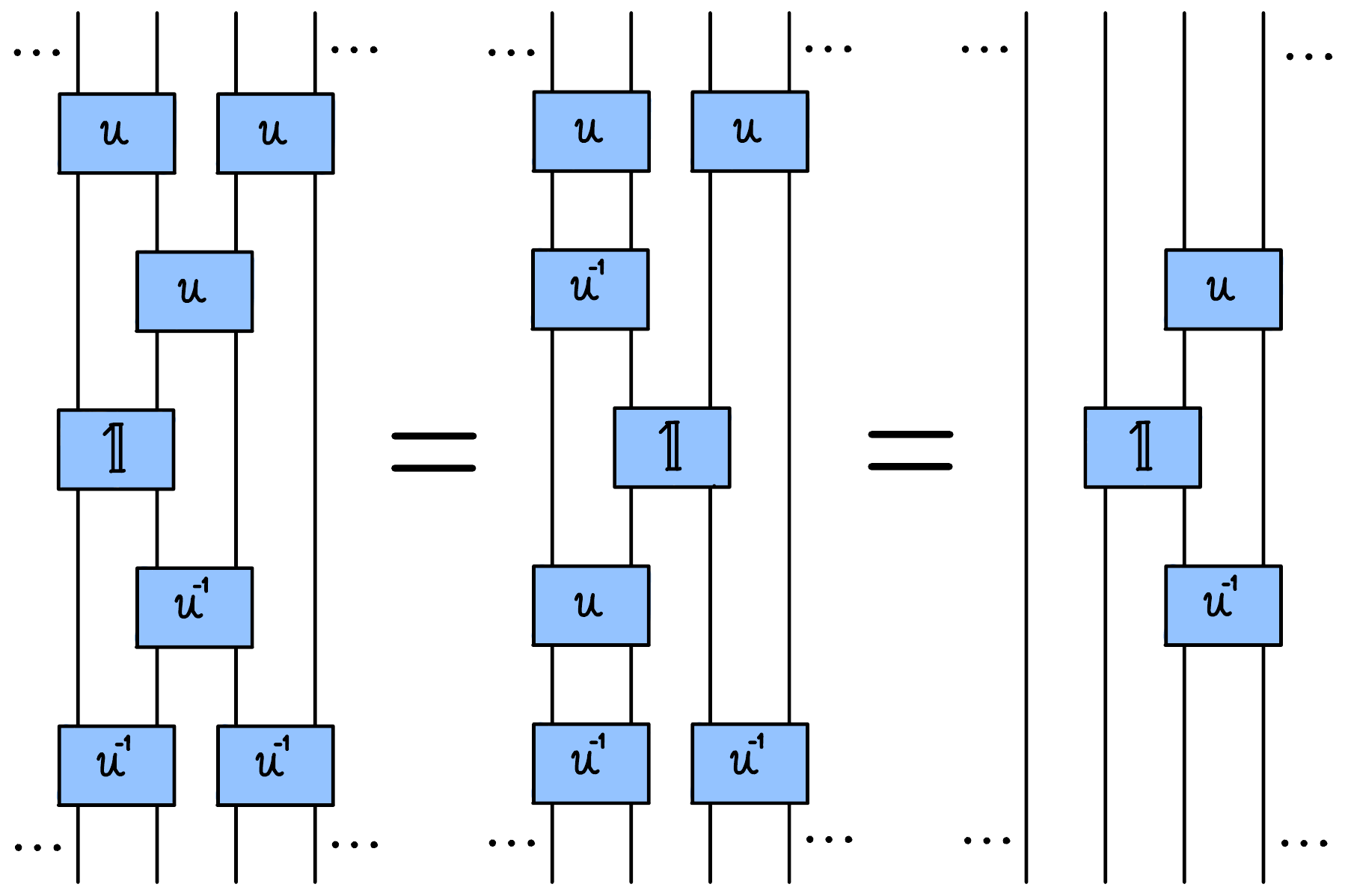}
\vspace{-1mm}
\caption{Diagrammatic proof of glider property (\ref{eq:glider1}) in terms of equivalence of quantum circuits, where gates
 $U[u],U[\one],U[u^{-1}]$ are indicated by blue boxes. The proof of (\ref{eq:glider2}) is analogous (via left-right reflection of the above diagrams).}
\label{fig:g}
\end{figure}

Consider a local Hilbert space $\mathcal H_1 = \mathbb C^3$ with states $\ket{s}_{s=0,+,-}$. We refer to states $\ket{\pm}$  as {\em charged particles} and we denote them by greek index $\ket{\alpha}_{\alpha=\pm}$, and to the state $\ket{0}$ as {\em vacancy}.
Let us define a local propagator over $\mathcal H_1 \otimes \mathcal H_1$ as
\bea
\!\!\!\!\!\!\!\!\!\!\!\!\!\!\!\!\!\!U[u] &=& \ket{00}\bra{00} + 
\sum_{\alpha\in\{\pm\}} (\ket{0\alpha}\bra{\alpha 0} + \ket{\alpha 0}\bra{0\alpha} +  \ket{\alpha\alpha}\bra{\alpha\alpha}) +  \\
\!\!\!\!\!\!\!\!\!\!\!\!\!\!\!\!\!\!&+& u_{11} \ket{+-}\bra{+-} + u_{12} \ket{+-}\bra{-+} +  u_{21} \ket{-+}\bra{+-} + u_{22} \ket{-+}\bra{-+}, \nonumber
\eea
where $u$ can be in principle any invertible matrix, $u\in GL(2)$, for the properties that will be discussed below, while for quantum physics applications we will think of $u$ as a unitary {\em mixing} matrix.
Such $U[u]$ satisfies two remarkable identities, valid for any $u$:
\bea
&& U_{12}[u] U_{23}[\one] U_{12}[u^{-1}] = U_{23}[u^{-1}]U_{12}[\one] U_{23}[u], \label{eq:id1}\\
&& U [u] U[u^{-1}] = \one, \label{eq:id2}
\eea
which can be checked by straightforward computation. Note that identity (\ref{eq:id1}) 
bears some similarity to the braid group form of Yang-Baxter equation if $u$ is considered as a non-abelian spectral parameter. 
Specifically,  $U[\one]$ obeys the braid relation -- it constitutes the so-called Yang-Baxter map -- and provides a deterministic update rule for the charged hardpoint lattice gas.
However $U[u]$ does not have the full Yang-Baxter property, but \emph{just enough} to render an existence of an infinite set of conserved local operators, as shown below.

For simplicity we can now assume a finite system of even number of sites $L$ and define its many-body Hilbert space as a tensor product $\mathcal H = \mathcal H_1^{\otimes L}$. Then we define a locality preserving discrete time dynamical system (a quantum cellular automaton) over it in terms of a brickwork circuit, i.e.
\be
\mathcal U = \mathcal U_{\rm o} \mathcal U_{\rm e},\quad
\mathcal U_{\rm e}  = \prod_{x=1}^{L/2} U_{2x-1,2x}[u],\quad \mathcal U_{\rm o} = \prod_{x=1}^{L/2} U_{2x,2x+1}[u], \quad L+1\equiv 1,
\ee
and write evolution of observables as
\be
a^{2t} = \mathcal U_{\rm e} a^{2t-1} \mathcal U_{\rm e}^{-1},\quad a^{2t+1} = \mathcal U_{\rm o} a^{2t} \mathcal U_{\rm o}^{-1}.
\ee
Let us define the following pair of local operators (note periodic boundary conditions if needed):
\bea
g^{(+)}_x &=& U_{x,x+1}[u]U_{x-1,x}U[\one]U_{x,x+1}[u^{-1}], \label{eq:glider}\\
g^{(-)}_x &=& U_{x-1,x}[u]U_{x,x+1}U[\one]U_{x-1,x}[u^{-1}]. \nonumber
\eea
Observing identities (\ref{eq:id1},\ref{eq:id2}), we can straightforwardly derive a few remarkable properties of these operators:
\bea
\mathcal U_{\rm e} g^{(+)}_{2x} \mathcal U^{-1}_{\rm e} &=& g^{(+)}_{2x+1},\qquad
\mathcal U_{\rm o} g^{(+)}_{2x-1} \mathcal U^{-1}_{\rm o} = g^{(+)}_{2x},\label{eq:glider1}\\
\mathcal U_{\rm e} g^{(-)}_{2x} \mathcal U^{-1}_{\rm e} &=& g^{(-)}_{2x-1},\qquad
\mathcal U_{\rm o} g^{(-)}_{2x+1} \mathcal U^{-1}_{\rm o} = g^{(-)}_{2x}, \label{eq:glider2}
\eea
which justify the term glider operators~\cite{gliders} (see Fig.~\ref{fig:g} for a simple graphical proof of the properties (\ref{eq:glider1},\ref{eq:glider2}) in terms of quantum circuit representation). 
We note that our gliders would be trivial $g^{(\pm)}_x=\one$ for {\em integrable trotterization}~\cite{zadnik} where $U(\one)=\one$.

The gliders are particular kind of quantum conservation laws, where the formal charge and current densities are identical.
Consequently, we can build products of such operators, separated by at least two pairs of sites, which again behave in the same way
\be
g^{(\pm)}_{x_1,x_2\ldots x_m} = \prod_{n=1}^m g^{(\pm)}_{2x_n},\quad \mathcal U g^{(\pm)}_{x_1,x_2\ldots x_m} \mathcal U^{-1} = g^{(\pm)}_{x_1\pm 1,x_2\pm 1\ldots x_m\pm 1}
\ee
under constraints $x_{j+1}-x_j \ge 2$.
We thus have a large set of extensive charges

\bea
Q^{(\pm)}_{r_1,r_2\ldots r_m} &=& \sum_{x=1}^{L/2} g^{(\pm)}_{x,x+r_1\ldots x+r_m}\,,\quad r_1\ge 2,\;r_{m+1}-r_m\ge 2,\nonumber \\ 
Q^{(\pm)}_{\un{r}} &\equiv& \mathcal U  Q^{({\pm})}_{\un{r}} 
 \mathcal U^{-1}.
\eea
These conserved operators concisely encapsulate the non-ergodic properties of the deformed hardpoint lattice gas.
The non-ergodicity should not be considered surprising as the pattern of left and right movers is evolving regularly according to ballistic (free) dynamics.
However, internal (charge) degrees of freedom undergo interacting dynamics which is nontrivially constrained by conserving $Q^{\pm}_{\un{r}}$.

The fact that the number of distinct glider charges $Q^{(\pm)}_{\un{r}}$ increases exponentially with the size of support, say $\ell$ such that all $r_j \le \ell$, is reminiscent of superintegrable quantum cellular automata~\cite{rule54review,balazsrule54,YBmaps}, yet the absence of exact Yang-Baxter
property suggests that the dynamical system is not completely integrable (i.e. there should be no Bethe ansatz!?). 
We are thus suggesting a new class of non-ergodic many-body quantum dynamics. 
It would be interesting to understand if there is any connection to a class of non-ergodic {\em dual unitary circuits} (for definition and basic properties see~\cite{BKP2019}) possessing glider operators without any 
(even partial) Yang-Baxter property (introduced and studied in Ref.~\cite{BP2022}).

\section{Conclusion}

The dynamical systems introduced and discussed in this short contribution have been constructed for the sole amusement of the author. Yet, we hope these constructions may 
at some point find some non-trivial applications in the context of non-equilibrium statistical mechanics. Specifically, one of the long-standing key goals in the field is to derive irreversible macroscopic transport laws, such as Fick's law of diffusion or Fourier's law of heat transport, from microscopic reversible equations of motion. 
This question has also been among the main research focuses of Giulio Casati~\cite{CasatiFourier} as well as stimulating the beginning of my research career~\cite{PR92}, and some of our joint work investigated the Fourier's law in systems very similar to the models studied here: hardpoint colliding masses in one dimension~\cite{CP2003}.

The coexistence of diffusive and ballistic transport has been rigorously demonstrated in Rule 54 
automaton (reviewed in Ref.~\cite{rule54review}, see also Ref.~\cite{sarang}), and one may hope that such results can be extended to the matter-field automaton, 
which seems richer and perhaps more generic (for example, Rule 54 dynamics is super-integrable with exponentially growing number of conserved local charges,  while the matter-field automaton is perhaps at most integrable as it has much rarer nontrivial local charges).

I acknowledge enjoying numerous discussions with Balazs Pozsgay on related problems.
This work has been supported by the Slovenian Research Agency (ARRS) under the Program P1-0402 and grants N1-0233 and N1-0219.

%%%%%%%%%%%%%%%%%%%%%%%%%%%%%%%%%%%%%%%%%%

 \end{document}